\documentclass[a4paper,twoside]{article}
\newcommand{\affil}[1]{$^{\rm #1}$}
\usepackage[authoryear]{natbib}
\bibpunct{(}{)}{;}{a}{}{,}
\usepackage{graphicx}
\usepackage{color}
\date{} 
\title{\large\bf\flushleft {\it s-} and {\it r-} process element abundances in the CMD of 47~Tucan\ae \ using the Robert Stobie Spectrograph on SALT$^1$}
\author{\parbox{\textwidth}{\flushleft
\vspace{-0.5cm}
{\it C.C.Worley\affil{A,C}, P.L.Cottrell\affil{A}, and E.C.Wylie~de~Boer\affil{B}}\\
\vspace{0.4cm}
{\small \affil{A}\,Dept of Physics \& Astronomy, University of Canterbury, Private Bag 4800, Christchurch, New Zealand}\\
{\small \affil{B}\,Research School of Astronomy \& Astrophysics, Mount Stromlo Observatory, Cotter Rd, Weston, ACT 2611, Australia}\\
{\small \affil{C}\,Email: ccw29@student.canterbury.ac.nz}}}
\begin{document}
\twocolumn[
\begin{changemargin}{.8cm}{.5cm}
\begin{minipage}{.9\textwidth}
\vspace{-1cm}
\maketitle
%
%
\small{\bf Abstract:}
A recent study by \citet{Wylie2006} has revealed that {\sl s-}process element abundances are enhanced relative to iron in both red giant branch and asymptotic giant branch stars of 47~Tucan\ae. A more detailed investigation into {\sl s-}process element abundances throughout the colour-magnitude diagram of 47~Tucan\ae \ is vital in order to determine whether the observed enhancements are intrinsic to the cluster. This paper explores this possibility through observational and theoretical means. The visibility of {\sl s-} and {\sl r-}process element lines in synthetic spectra of giant and dwarf stars throughout the colour magnitude diagram of 47~Tucan\ae \ has been explored. It was determined that a resolving power of 10 000 was sufficient to observe {\sl s-}process element abundance variations in globular cluster giant branch stars. These synthetic results were compared with the spectra of eleven 47~Tucan\ae \ giant branch stars observed during the performance verification of the Robert Stobie Spectrograph on the Southern African Large Telescope. Three {\sl s-}process elements, Zr, Ba, Nd, and one {\sl r-}process element, Eu, were investigated. No abundance variations were found such that {\footnotesize [X/Fe] = 0.0 $\pm$ 0.5} dex. It was concluded that this resolving power, $R \sim 5 \ 000$, was not sufficient to obtain exact abundances but upper limits on the {\sl s-}process element abundances could be determined. 


\vspace{0.2cm}
\small{ $^1$ Based on observations made with the Southern African Large Telescope (SALT)}

\vspace{0.2cm}
ACCEPTED December 2007: Publications of the Astronomical Society of Australia.

\medskip
\medskip
\end{minipage}
\end{changemargin}
]
\small

\section{{\normalsize {\bf Introduction}}}
Globular clusters are a rich area of study. They are very old and therefore can be studied as remnants of the early Galaxy and Universe. Their galactic orbits as individual clusters, and as a system of clusters, provide information regarding galaxy formation and structure. In addition, based on the assumption that all the stars in a globular cluster (GC) were formed from the same gas cloud, they provide a unique view of stellar evolution and nucleosynthesis. Variations in chemical abundances between the stars can be investigated as evidence of the different types of nuclear processing that occur as the stars evolve, or as evidence of inhomogeneities in the initial gas cloud. It is this latter area that the work presented here addresses.

Extensive literature is available on abundance variations, particularly carbon, nitrogen and oxygen abundances, within GCs \citep{Gratton2004}. As larger telescopes and more sensitive instruments become available, heavier elements have been added to the list of observed anomalies. Most recently, {\sl s-}process element enhancements have been observed in asymptotic giant branch (AGB) and red giant branch (RGB) stars in 47~Tucan\ae \ (47~Tuc) \citep{Wylie2006}. For AGB stars with mass greater than $1$ M$_{\odot}$ the third dredge-up (TDU) occurs with the mixing of {\sl s-}process elements formed in nuclear reactions in the C$^{13}$ pocket into the observable surface layers \citep{Busso2001}. However the AGB stars in 47~Tuc have mass less than $1$ M$_{\odot}$, making this an unlikely cause of the observed enhancements \citep{Gilliland1998}. There is no equivalent mechanism that would produce the observed enhancements in the RGB stars. The most likely explanation for the enhancements is that the gas cloud from which the stars formed was itself {\sl s-}process enhanced. Alternatively, the stellar atmospheres were polluted by stellar winds \citep{Cannon1998}.

Until relatively recently high resolution spectra could only be obtained for the brightest giants in GCs as the remaining stars were too faint for this type of observation. As the bright giants are evolved stars their atmospheres may have had the products of internal nuclear burning mixed through them, modifying their spectra. However, stars on the main sequence (MS) have not yet experienced mixing and it can be assumed that their spectra reflect the chemical composition of the initial gas cloud. 10~m class telescopes have the capabilities to observe these fainter dwarf stars in GCs. By comparing the spectra of main sequence stars to giant branch stars in the same cluster the intrinsic chemical composition of the cluster can be investigated. A new technique which pursues this comparison is to analyse the integrated light from a GC. This is a technique that is being developed and will also allow comparison of galactic to extra-galactic GCs \citep{McWilliam2007_accepted}.

The purpose of the current study has been to motivate a medium resolution study of 47~Tuc using the multi-object spectroscopy mode of the Robert Stobie Spectrograph on the Southern African Large Telescope and the AAOmega spectrograph on the Anglo-Australian Telescope.

\vspace{-0.25cm}
\section{{\normalsize {\bf Telescope $\&$ Instrument Combinations}}}
47~Tuc is the second largest observable galactic globular cluster and is only visible in the southern sky. Both the Southern African Large Telescope (SALT) and the Anglo-Australian Telescope (AAT) are ideally situated to perform a comprehensive survey of {\sl s-}process abundances in this cluster. 

SALT is a 10~m class telescope with instrumentation at the forefront of astronomical technology \citep{Buckley2006}. Currently in operation on SALT is the Robert Stobie Spectrograph (RSS) which has low to medium resolution capabilities \citep{Nordsieck2001}. The high resolution spectrograph (HRS) has completed the design stage \citep{Cottrell2005} and construction has commenced at the University of Durham in the UK.

AAT, a 3.9~m class telescope, has capabilities for multiple object surveys using AAOmega and 2dF \citep{Sharp2006}. These instruments have already been used successfully for element abundance surveys of globular clusters, including 47~Tuc \citep{Cannon2003}.

This paper considers the feasibility of using RSS and AAOmega for the 47~Tuc {\sl s-}process abundance survey. These instruments have comparable maximum resolving powers, RSS at $R \sim 10 \ 000$ and AAOmega at $R \sim 8 \ 000$. Comparison will also be made with the high resolving power capabilities of SALT HRS, which has a nominal maximum resolving power of $R \sim   60 \ 000$. Section~\ref{LSA} addresses the question of whether or not variations in {\sl s-}process abundances of at least $+0.5$ dex will be visible in spectra obtained at the maximum operating resolving power of these instruments. The line strength visibility for a range of key {\sl r-} and {\sl s-}process species will be explored in the effective temperature ($T_{\textrm{\scriptsize{eff}}}$) - surface gravity ($log \ g$) space represented by the colour magnitude diagram (CMD) of 47~Tuc for both high and medium resolving powers.
Section~\ref{SPV} presents the results of abundance analyses of {\sl s-}process elements in eleven giant branch stars in 47~Tuc that were observed using RSS on SALT in 2006.

\vspace{-0.25cm}
\section{{\normalsize {\bf Stellar Atmospheres $\&$ Line Strength}}}\label{LSA}
The chemical composition of a star is determined by the analysis of absorption lines in the stellar spectrum. Every line corresponds to the absorption of energy by an atom and there is a direct relation between the strength of a line and how much of the corresponding element is present in the star's atmosphere. However there are other factors which affect the observed strength of a line. These include the temperature, pressure and opacity in the star's atmosphere. When considering the stars in a globular cluster the two key parameters for classifying the evolutionary stage of each star are the effective temperature, $T_{\textrm{\scriptsize{eff}}}$, and the gravity, $log \ g$. 

\subsection{{\normalsize {\bf Line Strength Analysis in the 47~Tuc `$T_{\textrm{\scriptsize{eff}}} - log \ g$' Space}}}
\begin{figure}[h]
\begin{center}
\caption{{\footnotesize CMD of 47~Tuc \citep{Cannon1998,Lee1977,Hartwick1974} showing $T_{\textrm{\scriptsize{eff}}}$ and $log \ g$ ($\circ$) at which synthesised spectra are computed and stars observed using SALT RSS during performance verification ($\star$).}}\label{models}
\includegraphics[width=80mm]{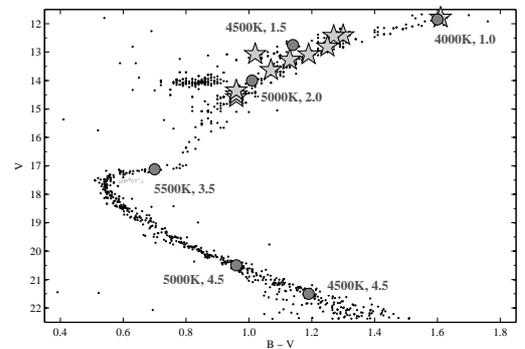}
\end{center}
\end{figure}

Figure~\ref{models} is the CMD of 47~Tuc \citep{Cannon1998,Lee1977,Hartwick1974}. The stars range in $T_{\textrm{\scriptsize{eff}}}$ from $\sim 4000$~K to $\sim 6000$~K, and in $log \ g$ from $\sim 1.0$ to $\sim 4.5$.

\vspace{-0.2cm}

\vspace{-0.5cm}
The six circular points on the CMD shown in Figure~\ref{models} were selected to represent key stages of stellar evolution at which {\sl s-}process element abundance analysis would be undertaken. The $T_{\textrm{\scriptsize{eff}}}$ and $log \ g$ values for each point are displayed. The eleven star-shaped points are the locations of the eleven 47~Tuc stars observed during the performance verification of RSS on SALT which will be discussed in Section~\ref{SPV}.

The models and linelists used in the spectrum synthesis programme ({\footnotesize MOOG},\citet{MOOG}) were sourced from the Kurucz website ({\footnotesize http://kurucz.harvard.edu}). Each model had a metallicity of {\footnotesize [Fe/H] = -0.5} which is close to the metallicity of 47~Tuc ({\footnotesize [Fe/H] = -0.7}) \citep{Harris1996}. A microturbulence of $2.0$~kms$^{-1}$ was selected. The Kurucz linelists are semi-empirical and some adjustment of oscillator strength ($log \ gf$) was required for individual lines \citep{Kurucz1975}. In particular, they were adjusted to include recent laboratory $log \ gf$ values for {\sl r-} and {\sl s-}process elements \citep{Biemont1981, Den2003, Hannaford1982, Lawler2001}. Initially the linelists and models were used to produce spectra that were compared to the high resolution atlas of Arcturus which is available on the NOAO website (\small{http://www.archive.noao.edu}). Arcturus has a similar metallicity to 47~Tuc stars. The $log \ gf$ values of other species in the linelists were adjusted to provide a best fit to Arcturus.

Synthetic spectra were computed for each identified evolutionary point in Figure~\ref{models}. Three spectral regions were synthesised based on lines used in \citet{Wylie2006}. The key species present in each of the spectral regions are set out in Table~\ref{spec_reg}, where light {\sl s-} represents the elements Y, Sr and Zr, and heavy {\sl s-} represents Nd, La and Ba.

\vspace{-0.25cm}
\begin{table}[h]
\begin{center}
\caption{{\footnotesize Key features in synthesised regions.}}\label{spec_reg}
\vspace{0.2cm}
{\small
\begin{tabular}{cccc} 
\hline Region ({\AA}) & Feature ({\AA}) & Species & Process \\ 
\hline 6139 - 6145 & 6140.46 & Zr~I & light {\sl s-} \\ 
  & 6143.18 & Zr~I & light {\sl s-} \\ 
5803 - 5807 & 5804.00 & Nd~II & heavy {\sl s-} \\ 
  & 5805.77 & La~II & heavy {\sl s-} \\ 
6644 - 6646 & 6645.13 & Eu~II & {\sl r-} \\ 
\hline
\end{tabular}
}
\end{center}
\end{table}

\vspace{-0.25cm}
Model spectra were synthesised at high ($R \sim 60 \ 000$) and medium ($R \sim 10 \ 000$) resolving powers. The high resolving power provides comparison with a current instrument (UCLES) at the AAT \citep{Wylie2006} and an equivalent instrument, SALT HRS, currently under contruction \citep{Buckley2006}. The medium resolving power is about the highest achievable with RSS on SALT and comparable to AAOmega's maximum resolving power on the AAT.

Each of Figures~\ref{m13} to~\ref{m5} and Figures~\ref{m8} to~\ref{m4} corresponds to one of the six stellar models shown in the CMD of 47~Tuc in Figure~\ref{models}. In each spectral region the abundance of the key species was varied with respect to the model abundance.

By comparing the spectral regions between the models, and between the two resolving powers, the changes in line strength for the different species are very clear.

\subsection{{\normalsize {\bf $T_{\textrm{\scriptsize{eff}}}$ \& $log \ g$ effects on the GB}}}
\begin{figure}[h!]
\begin{center}
\caption{{\footnotesize Synthetic giant branch spectra for key features computed at $T_{\textrm{\scriptsize{eff}}} = 4000$~K, $log \ g = 1.0$. The dash-dot line represents a variation of $-8.0$ dex simulating when none of the species, X, is present. The solid line represents no variation of the species abundance from the model {\footnotesize ([X/Fe]=0.0)}. The dashed line represents a variation of $+0.5$ dex and the dotted line, $+1.0$ dex.}}\label{m13}
\includegraphics[width = 80mm,height=53mm]{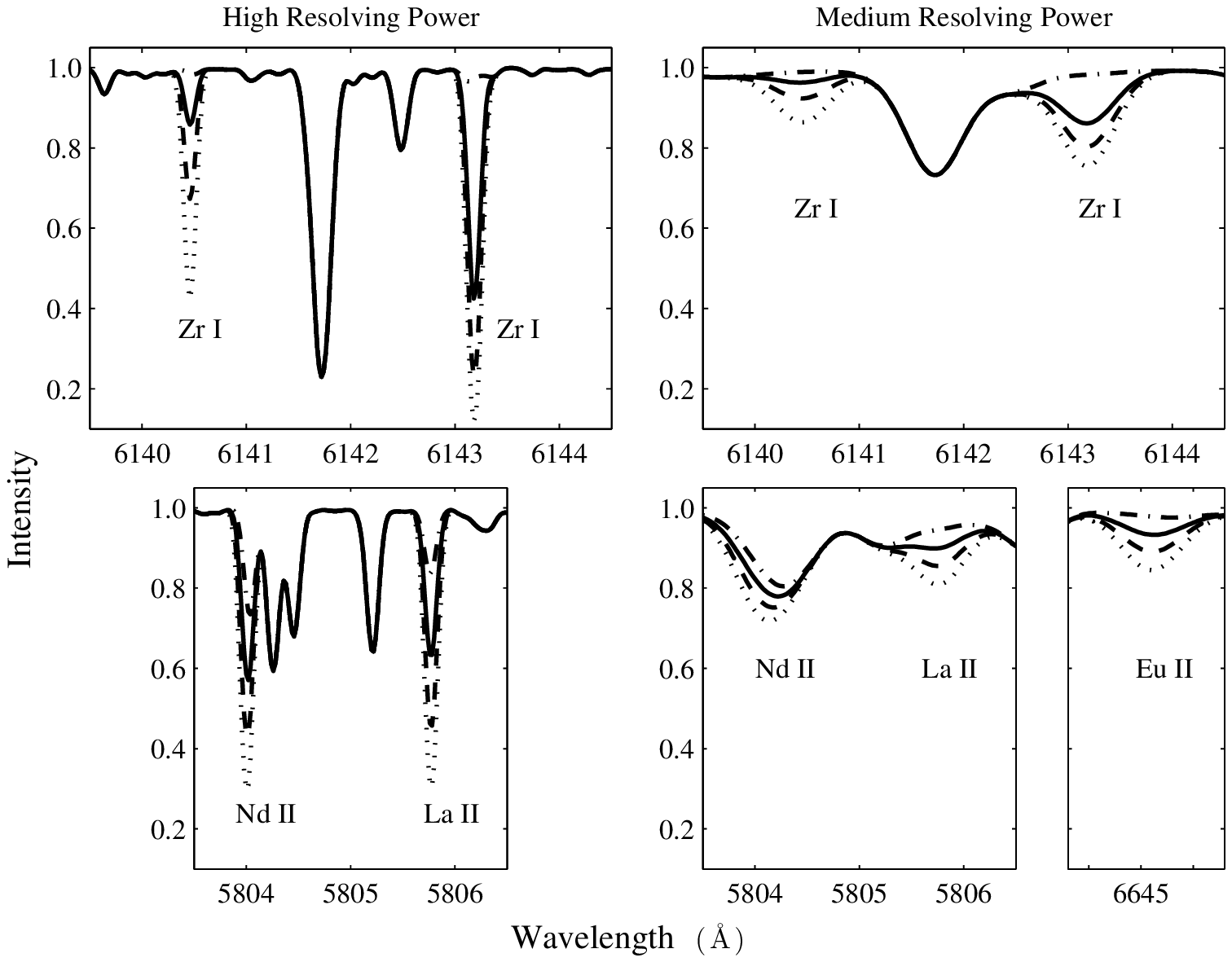}
\caption{{\footnotesize As for Figure~\ref{m13} but $T_{\textrm{\scriptsize{eff}}} = 4500$~K, $log \ g = 1.5$.}}\label{m3}
\includegraphics[width =
    80mm,height=53mm]{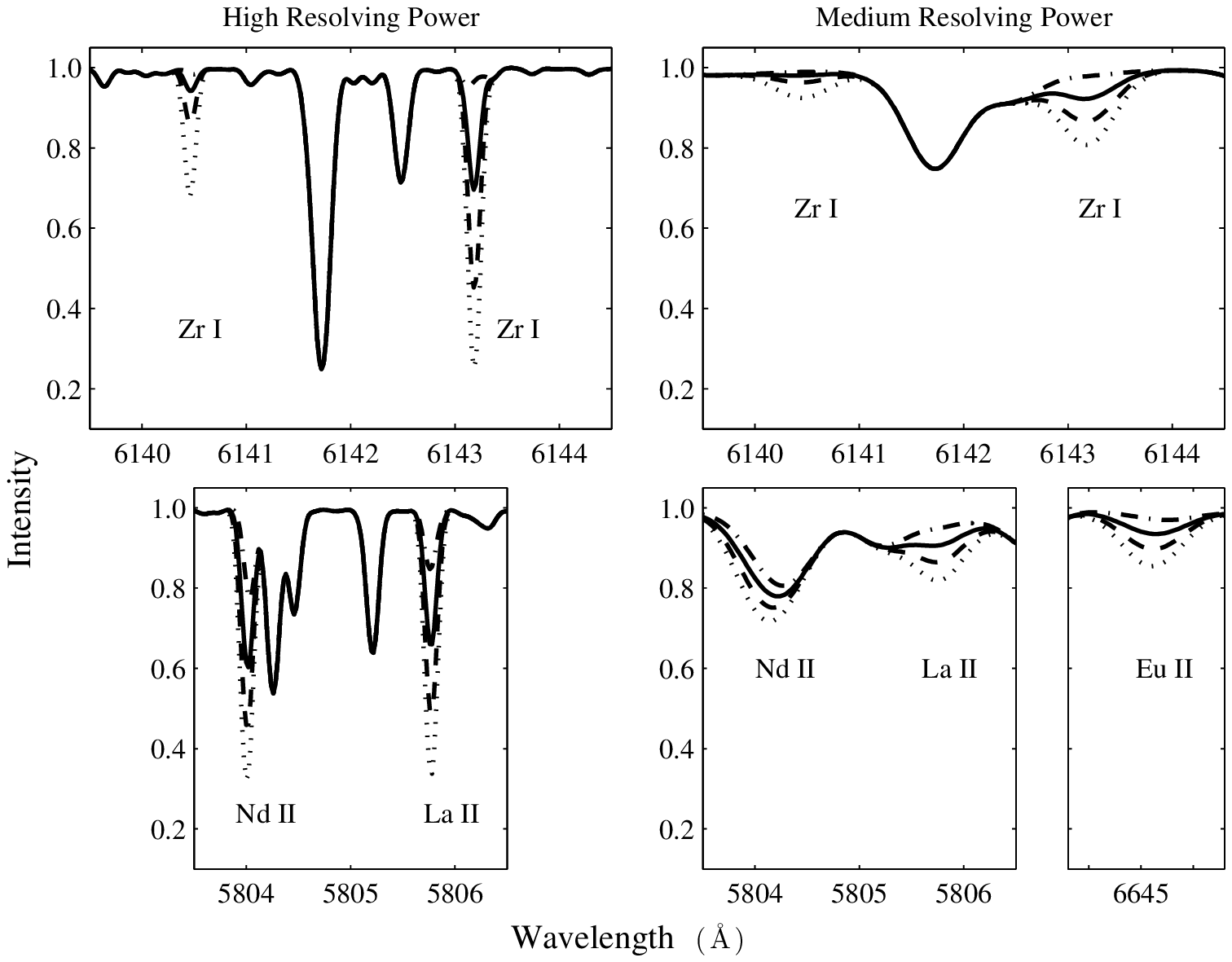}
\caption{{\footnotesize As for Figure~\ref{m13} but $T_{\textrm{\scriptsize{eff}}} = 5000$~K, $log \ g = 2.0$.}}\label{m5}
\includegraphics[width =
    80mm,height=53mm]{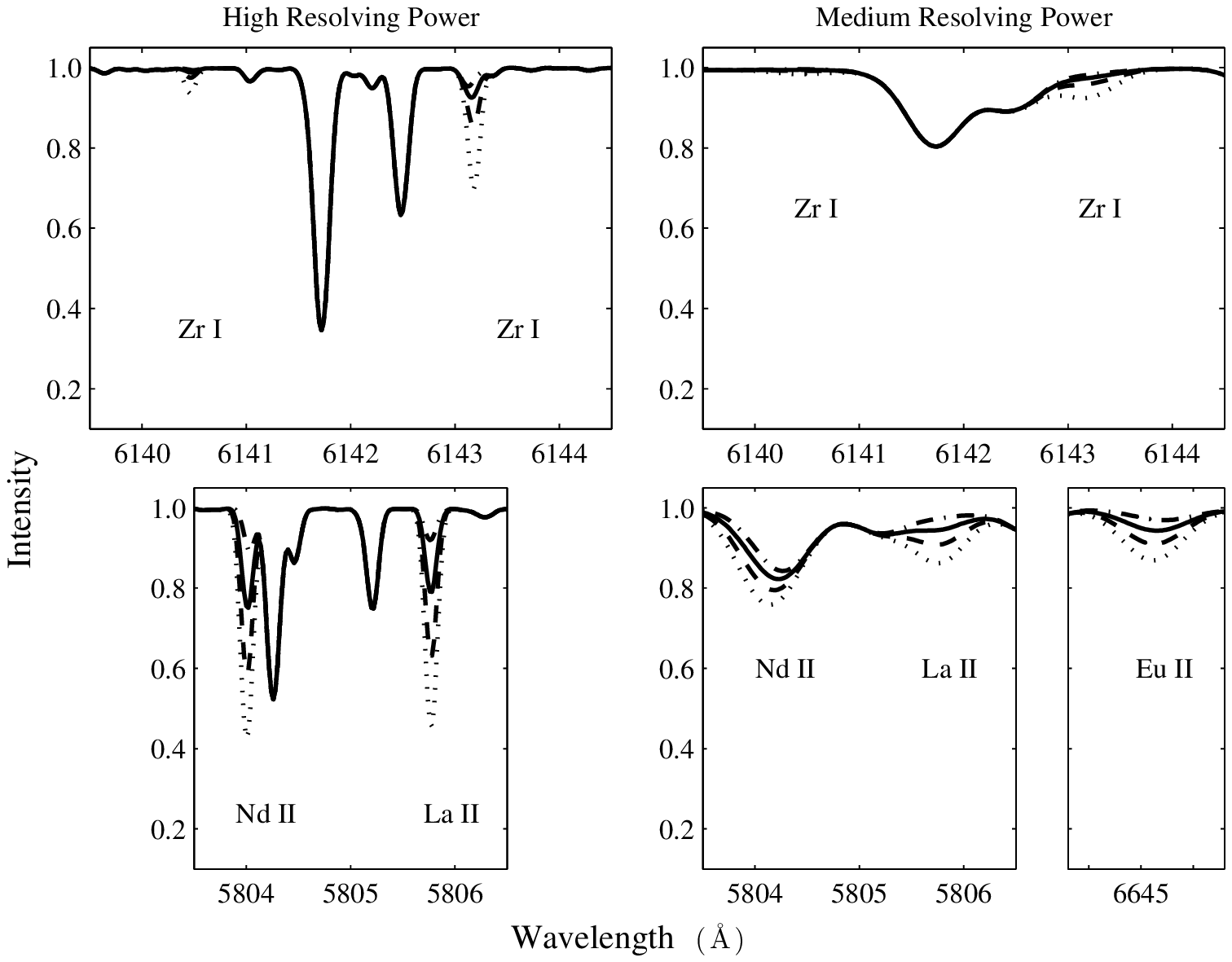}
\end{center}
\end{figure}

Figures~\ref{m13},~\ref{m3} and~\ref{m5} show the synthesised spectra for the three giant branch evolutionary stages. The progression from the tip down the giant branch involves an increase in both $T_{\textrm{\scriptsize{eff}}}$ and $log \ g$. 

Both the neutral and ionised lines show a decrease in line strength down the giant branch. However the ionised lines are much less affected than the neutral lines. This is expected as the ionised state becomes dominant due to the increasing temperature. However the increased temperature also increases the  $H^{-}$ continuous opacity which weakens both line species. 

The increased pressure has more complicated effects as the neutral dominated populations are changing to ionised dominated populations due to the increased temperature. Initially the neutral lines will weaken with the increase in pressure, but as the population becomes ionised the effect becomes negligible. The ionised lines will initially weaken considerably with the increased pressure, but as the population becomes ionised this effect will lessen. 

The key observable in this sequence is the weakening of Zr~I lines as the temperature and pressure increase, while the La~II, Nd~II and Eu~II lines are less affected. 
The medium resolving power synthesis replicates the changes in line strength outlined for the high resolving power. All the features are identifiable and distinguishable at the +0.5 dex level, although the Zr~I lines at the highest temperature have weakened considerably and would not be of use as Zr abundance indicators at medium resolving power.

\section{{\normalsize {\bf SALT Performance Verification}}}\label{SPV}
In June and October of 2006 spectra for eleven giant branch stars in 47 Tuc were obtained during the performance verification commissioning stage of RSS on SALT. Figure~\ref{models} shows the location of the eleven stars on the 47 Tuc CMD. The observations were carried out at two different camera settings obtaining two overlapping wavelength regions: 6020-6860~{\AA} \& 5200-6150~{\AA}.

These regions contain the key {\sl s-} and {\sl r-}process lines that were used in the study of theoretical line strength. The resolving power obtained for these spectra was $\sim 5000$, which is not the maximum that will be possible with RSS. Signal to noise ratios between 50 and 140 were obtained.

Table~\ref{tempgrav} presents the effective temperatures and surface gravities for each star that were calculated from the stellar apparent magnitudes and colour indices using the calibration equations from \citet{Alonso1999}.

\vspace{-0.5cm}
\begin{table}[h]
\begin{center}
\caption{{\footnotesize $T_{\textrm{\scriptsize{eff}}}$ and $log \ g$ values from \citet{Alonso1999} for each of the SALT PV Stars.}}\label{tempgrav}
\vspace{0.2cm}
{\small
\begin{tabular}{ccccc} 
\hline  Lee No. & V & B-V & $T_{\textrm{\scriptsize{eff}}}$ (K) & $log \ g$ \\ 
\hline 3512 & 11.79 & 1.63 & 4080 & 0.7 \\ 
1513 & 12.41 & 1.32 & 4450 & 1.3 \\ 
2525 & 12.43 & 1.29 & 4490 & 1.4 \\
6519 & 12.81 & 1.27 & 4520 & 1.5 \\
6524 & 13.08 & 1.21 & 4610 & 1.7 \\
1506 & 13.27 & 1.15 & 4700 & 1.9 \\
3510 & 13.63 & 1.09 & 4800 & 2.1 \\
2604 & 13.07 & 1.04 & 4890 & 1.9 \\
4514 & 14.35 & 0.98 & 5000 & 2.5 \\
4515 & 14.49 & 0.98 & 5000 & 2.5 \\
4513 & 14.59 & 0.98 & 5000 & 2.6 \\
\hline
\end{tabular}
}
\end{center}
\end{table}

\vspace{-0.5cm}
\subsection{{\normalsize {\bf Temperature Sensitivity}}}
\begin{figure}[!h]
\caption{{\footnotesize Temperature sensitivity: Zr I \& Ba II. Key regions for observed (left panel) \& synthesised (right panel) spectra. The key for the synthesised spectra as per Figure~\ref{m13}.}}\label{zrtemp}
\includegraphics[width = 81mm]{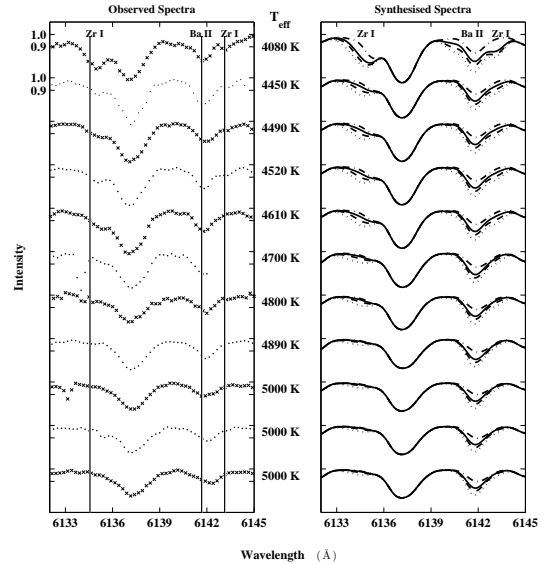}
\end{figure}

Figure~\ref{zrtemp} shows both the observed spectra of the eleven stars and the synthesised spectra of the region containing key Zr I  and Ba II features.

The spectra are presented with the temperature increasing down the panel, replicating the traverse down the giant branch. This region contains two Zr I features and one Ba II feature. The vertical lines trace the line locations in each spectrum.  

In the coolest stars the neutral Zr features are quite distinct and get weaker until a temperature of around 5000 K. The ionised Ba, while it does weaken slightly, remains apparent through to the hotter temperatures. 

This nicely replicates the theoretical analysis, showing the more dramatic weakening of neutral lines compared with ionised lines. It also is a good illustration of temperature sensitivity, the strength of the Zr I features providing a temperature scale that confirms the scale derived from the \citet{Alonso1999} equations. The temperature sensitivity is replicated in the synthetic spectra. 

\subsection{{\normalsize {\bf Abundance Results}}}
\begin{figure}[h!]
\begin{center}
\caption{{\footnotesize Abundances: Zr I \& Ba II. As for Figure~\ref{zrtemp} but observed spectrum overlaid on synthesised spectra.}}\label{zrabun}
\includegraphics[width = 75mm]{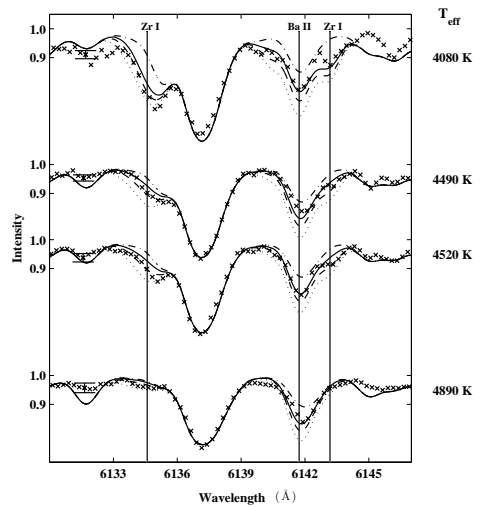}
\end{center}
\end{figure}

\begin{figure}[h!]
\begin{center}
\caption{{\footnotesize Abundances: Nd II \& Eu II. Key as for Figure~\ref{zrtemp} but observed spectrum overlaid on synthesised spectra. The observed spectra with $T_{\textrm{\scriptsize{eff}}} = 4890$~K shows an absorption artifact due to a cosmic ray.}}\label{ndabun}
\includegraphics[width = 75mm]{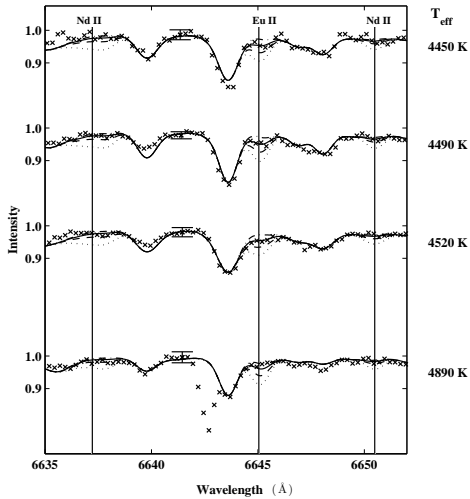}
\end{center}
\end{figure}

Figures~\ref{zrabun} and~\ref{ndabun} present an abundance analysis of the two key spectral regions. In Figure~\ref{zrabun} the synthetic spectra (as for Figure~\ref{m13}) are overlaid onto the observed spectra for the Zr I \& Ba II region for four of the stars. The error bars give an indication of the uncertainty due to noise in these spectra. 

\vspace{-0.5cm}

\vspace{-0.5cm}
The coolest star provides the best estimate. The Zr I features are prominent, again traced out by the vertical line. This spectrum best fits the model synthesis, indicating no Zr enhancement. The Ba II feature was also fitted best by the model abundance synthesis.

For the next two spectra, both of $\sim 4500$~K, while within error {\footnotesize [Ba/Fe], [Zr/Fe]} = 0.0 there is the possibility of some Zr enhancement at a +0.5 dex level.

For the hotter stars the Zr I features have diminished too much to observe abundance variations. The Ba II feature is still prominent but the noise levels of these spectra do not give a convincing argument for anything other than model abundance.
Overall an {\footnotesize [Zr/Fe]} = 0.0 with a possible uncertainty to + 0.5dex, while {\footnotesize [Ba/Fe]} = 0.0 for all stars.

Figure~\ref{ndabun} presents a similar abundance analysis of the region containing key Nd II \& Eu II features. Even for the coolest star the Nd II features are too weak to provide a reliable analysis. The Eu II feature is more promising for the cooler stars (4500 K) with a good fit to the model abundance ({\footnotesize [Eu/Fe]} = 0.0) within error. As the stars get hotter this feature also diminishes. 

\vspace{-0.5cm}

\vspace{-0.5cm}
The analysis of these spectra shows that {\footnotesize [X/Fe]}~=~0.0 is the best solution in all cases with upper limits on the abundance values of +0.5~dex. For future RSS observations obtaining the expected maximum resolving power of $10 \ 000$ will provide the required resolution to refine this upper limit.

\vspace{-0.25cm}
\section{{\normalsize {\bf Future work - Main Sequence}}}
Comparison between the AGB and the RGB of a GC will provide a test for TDU if the AGB stars are {\sl s-}process enhanced while the RGB stars are not. However, as the spectra of GC MS stars correspond to the chemical composition of the initial GC gas cloud, to really investigate pollution versus nucleosynthesis theories the investigation must inevitably be extended to include the MS. The 10 m class of telescopes have the capability of observing these faint dwarf stars in distant GCs. As discussed earlier, this investigation is also being pursued via integrated light techniques on 2.5 m class telescopes \citep{McWilliam2007_accepted}.

\begin{figure}[ht]
\begin{center}
\caption{{\footnotesize $V$ magnitude limits on 47~Tuc CMD with current (solid lines) and future (dashed lines) observations on SALT and the AAT.}}\label{vis_lim}
\includegraphics[width = 70mm,height=55mm]{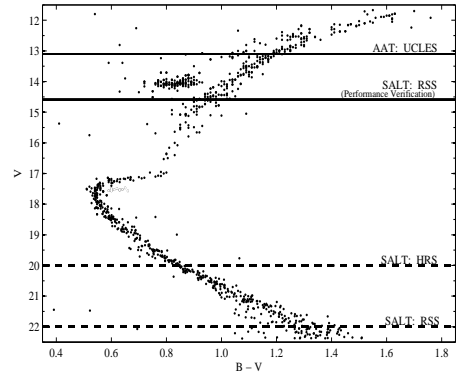}
\end{center}
\end{figure}

Figure~\ref{vis_lim} shows the apparent magnitude limits of current (solid line) and future (dashed line) samples of 47~Tuc stars from high and medium resolution instruments on SALT and the AAT. The SALT RSS limiting magnitude is expected to be $\sim 22$~mag which is sufficient for the spectra of main sequence stars in 47~Tuc to be obtained.

\vspace{-0.5cm}
\subsection{{\normalsize {\bf $T_{\textrm{\scriptsize{eff}}}$ \& $log \ g$ effects on the MS}}}
\begin{figure}[h!]
\begin{center}
\caption{{\footnotesize Synthetic Spectra for key features computed at  $T_{\textrm{\scriptsize{eff}}} = 5500$~K, $log \ g = 3.5$ for a main sequence turn off star with key as for Figure~\ref{m13}.}}\label{m8}
\includegraphics[width =
    80mm,height=53mm]{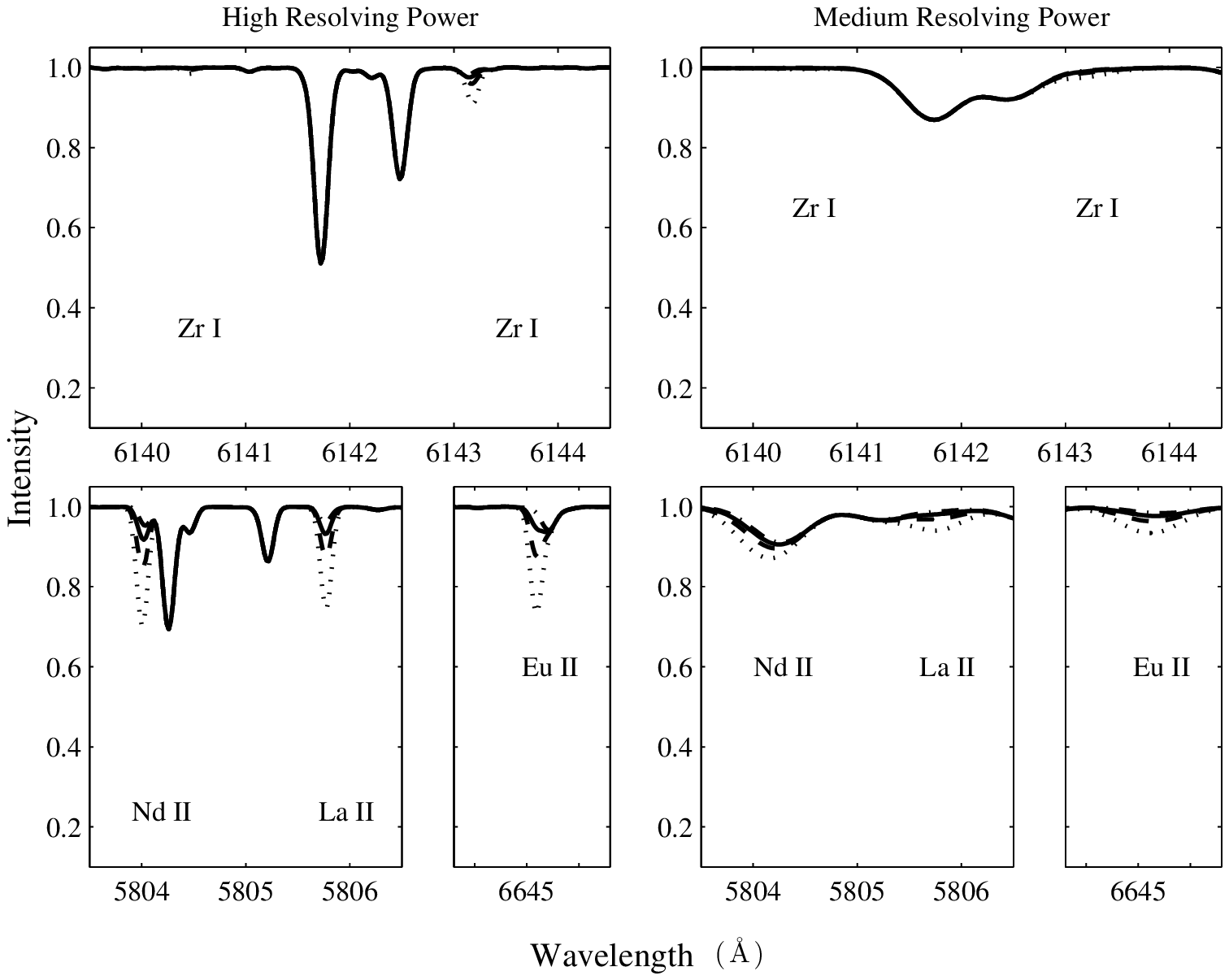}
\caption{{\footnotesize As for Figure~\ref{m8}, except on the main sequence at $T_{\textrm{\scriptsize{eff}}} = 5000$~K, $log \ g = 4.5$.}}\label{m6}
\includegraphics[width =
    80mm,height=53mm]{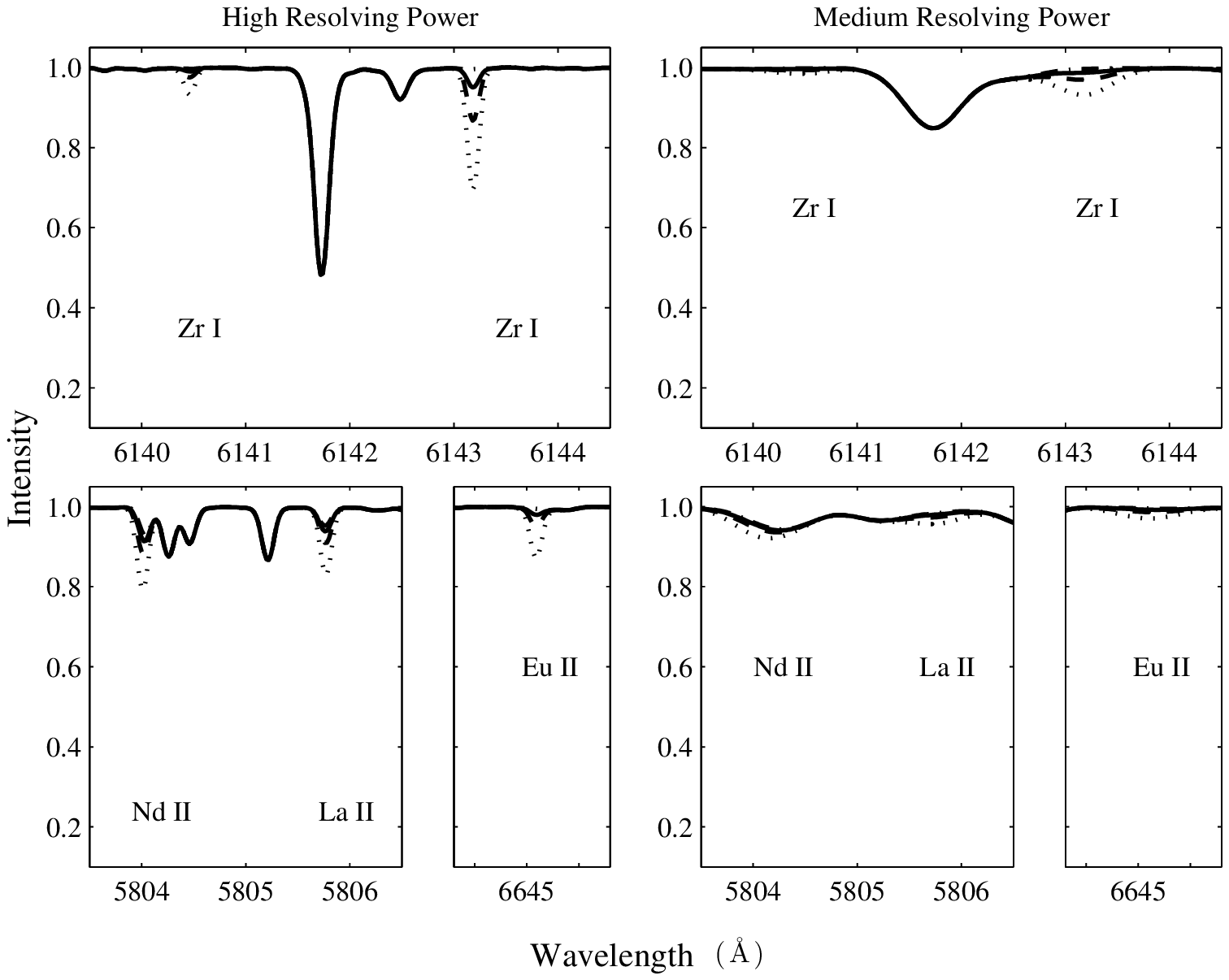}
\caption{{\footnotesize As for Figure~\ref{m8}, except on main sequence at  $T_{\textrm{\scriptsize{eff}}} = 4500$~K, $log \ g = 4.5.$}}\label{m4}
\includegraphics[width =
    80mm,height=53mm]{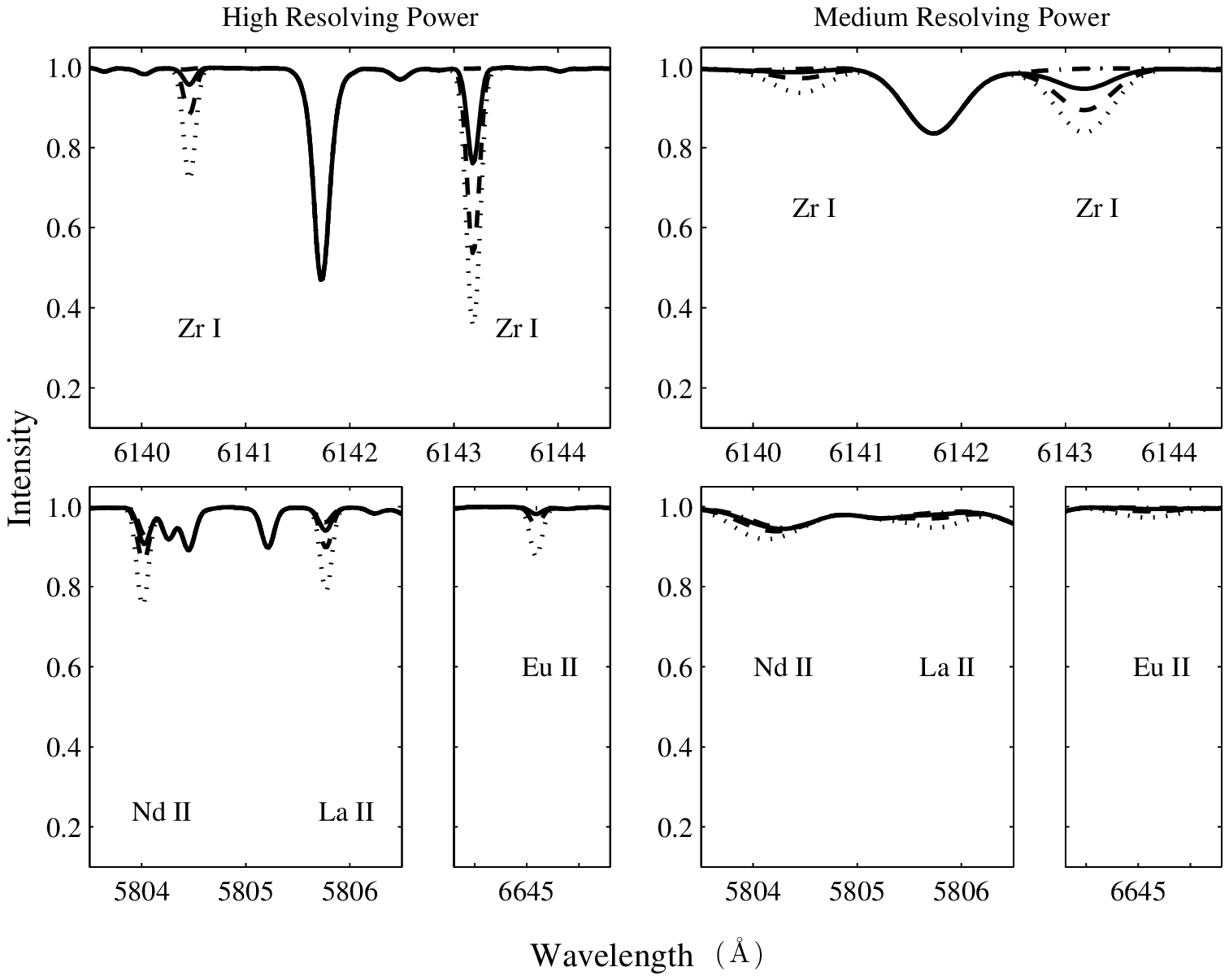}
\end{center}
\end{figure}

Of the six evolutionary points on the 47~Tuc CMD in Figure~\ref{models}, one lies on the main sequence turn off, and two lie on the MS. Synthetic spectra were generated for these points as for the GB points in order to determine the visibility of {\sl s-} and {\sl r-}process lines at these temperatures and gravities. These spectra are shown in Figures~\ref{m8} to~\ref{m4}. 

Progressing from the giant branch, the increased temperature causes increased ionisation which strengthens the ionised lines. However the lines are weakened due to the increase in continuous opacity and the increased pressure to which ionised lines are sensitive. Figure~\ref{m8} shows that the line strength of the ionised lines has reduced significantly in the high and medium resolving power spectra.

The effect on the neutral lines is much more dramatic. Both the increased ionisation and increased continuous opacity weaken the neutral lines but there is no effect at work to strengthen the neutral lines. Figure~\ref{m8} shows that in the high resolving power spectra the neutral lines have almost disappeared while in the medium resolving power spectra they are non-existent.

Figures~\ref{m6} and~\ref{m4} show the synthesised spectra for the two points on the MS. The models have the same $log \ g$ value, so there is no change in pressure, but the $T_{\textrm{\scriptsize{eff}}}$ is lower for the point further down the MS. The differences in line strength between the models is due to the decrease in temperature. However the pressure is very high and the change from an ionised to neutral dominated population increases the pressure sensitivity of both the ionised lines and neutral lines.

The neutral lines start to experience the weakening effects of the high pressure as the population returns to being neutral dominated. However the decreased temperature means decreased ionisation and a decrease in continuous opacity.

\vspace{-0.35cm}
 These act to significantly strengthen the neutral lines on the descent down the MS (see Figures~\ref{m6} and ~\ref{m4}). The Zr~I features have become very pronounced in the lowest temperature model. For the ionised lines the comparison between Figures~\ref{m6} and~\ref{m4} shows that overall there is some strengthening of the ionised lines due to the decrease in temperature but the strong pressure ensures it is a small effect. 

This comparison indicates that neutral lines are the best candidates for abundance analysis of MS stars in both the high and medium resolving powers. The resolving power of SALT HRS ($R \sim 60 \ 000$) will be required to fully test the hypothesis of intrinsic MS enhancements using these features. However those MS stars above RSS's limiting magnitude but still sufficiently low in temperature will have neutral lines that are strong enough to be observed at medium resolving power.

\section{{\normalsize {\bf Conclusions}}}
The analysis of these initial (PV) SALT RSS observations of eleven stars in 47~Tuc indicate no enhancement of Zr, Ba or Eu, that is {\footnotesize [X/Fe]} = 0.0 for these three species. However, the uncertainty is sufficiently large (+0.5 dex) that there may be some star-to-star variation. This will have to await the more extensive survey at the maximum resolving power ($\sim 10 \ 000$) of RSS. The SALT RSS PV observations do provide an excellent temperature sequence and confirmation of the effective temperatures derived from \citet{Alonso1999} calibrations.

The exploration of the $T_{\textrm{\scriptsize{eff}}} - log \ g$ space in terms of line strength has significantly refined the line selection process for the abundance survey. The analysis shows that neutral lines are observable over a greater range of $log \ g$ values than the ionised lines, but within a smaller range of $T_{\textrm{\scriptsize{eff}}}$ values. Conversely the ionised lines are observable over a greater range of $T_{\textrm{\scriptsize{eff}}}$ values than the neutral lines, but within a smaller range of $log \ g$ values. This broadens the range of temperatures and gravities from which the survey star sample will be selected.

The characteristics of the instruments under consideration further limit the star sample. The medium resolving power of RSS and the magnitude limit of SALT do reduce the sample options with regards to the MS. There is a small range on the MS where the temperatures are cool enough and the magnitudes are bright enough that abundances could be obtained using RSS, though SALT HRS is more suited to obtaining this data. However, {\sl s-}process element abundance analysis to just below the horizontal branch, which this analysis shows is quite possible using RSS or AAOmega, will provide definitive results on the nature of {\sl s-}process element abundance variations in 47~Tuc.

The next stage of this investigation is to consider globular clusters over a range of metallicity. The stellar models considered so far have a metallicity similar to that of 47~Tuc. The {\sl s-}process element abundance survey results for 47~Tuc will then need to be compared to observational surveys of other clusters at different metallicity. This will show whether any {\sl s-}process element abundance variations are isolated to 47~Tuc, or are a general characteristic of globular clusters.

\section*{Acknowledgements}
This research has been supported by the University of Canterbury Scholarship programme, the Dennis William Moore Scholarship Fund and a 2007 Fellowship from the New Zealand Federation of Graduate Women.

The observations reported in this paper were obtained with the Southern African Large Telescope (SALT), a consortium consisting of the National Research Foundation of South Africa,
Nicholas Copernicus Astronomical Center of the Polish Academy of Sciences, Hobby Eberly Telescope Founding Institutions, Rutgers University, Georg-August-Universitat Gottingen, University of Wisconsin-Madison, Carnegie Mellon University, University of Canterbury, United Kingdom SALT Consortium, University of North Carolina - Chapel Hill, Dartmouth College, American Museum of Natural History and the Inter-University Centre for Astronomy and Astrophysics, India.


\begin{thebibliography}{21}
\expandafter\ifx\csname natexlab\endcsname\relax\def\natexlab#1{#1}\fi

\bibitem[{{Alonso} {et~al.}(1999){Alonso}, {Arribas}, \&
  {Mart{\'{\i}}nez-Roger}}]{Alonso1999}
{Alonso}, A., {Arribas}, S., \& {Mart{\'{\i}}nez-Roger}, C. 1999, A$\&$AS, 140,
  261

\bibitem[{{Biemont} {et~al.}(1981){Biemont}, {Grevesse}, {Hannaford}, \&
  {Lowe}}]{Biemont1981}
{Biemont}, E., {Grevesse}, N., {Hannaford}, P., \& {Lowe}, R.~M. 1981, ApJ,
  248, 867

\bibitem[{{Buckley} {et~al.}(2006){Buckley}, {Burgh}, {Cottrell}, {Nordsieck},
  {O'Donoghue}, \& {Williams}}]{Buckley2006}
{Buckley}, D.~A.~H., {Burgh}, E.~B., {Cottrell}, P.~L., {Nordsieck}, K.~H.,
  {O'Donoghue}, D., \& {Williams}, T.~B. 2006, in {Proceedings of SPIE}, Vol.
  6269, {Ground-based and Airborne Instrumentation for Astronomy.}, ed. I.~S.
  {McLean} \& I.~{Masanori}, 62690A

\bibitem[{{Busso} {et~al.}(2001){Busso}, {Gallino}, {Lambert}, {Travaglio}, \&
  {Smith}}]{Busso2001}
{Busso}, M., {Gallino}, R., {Lambert}, D.~L., {Travaglio}, C., \& {Smith},
  V.~V. 2001, ApJ, 557, 802

\bibitem[{{Cannon} {et~al.}(2003){Cannon}, {da Costa}, {Norris}, {Stanford}, \&
  {Croke}}]{Cannon2003}
{Cannon}, R., {da Costa}, G., {Norris}, J., {Stanford}, L., \& {Croke}, B.
  2003, in Astronomical Society of the Pacific Conference Series, Vol. 296, New
  Horizons in Globular Cluster Astronomy, ed. G.~{Piotto}, G.~{Meylan}, S.~G.
  {Djorgovski}, \& M.~{Riello}, 175

\bibitem[{{Cannon} {et~al.}(1998){Cannon}, {Croke}, {Bell}, {Hesser}, \&
  {Stathakis}}]{Cannon1998}
{Cannon}, R.~D., {Croke}, B.~F.~W., {Bell}, R.~A., {Hesser}, J.~E., \&
  {Stathakis}, R.~A. 1998, MNRAS, 298, 601

\bibitem[{{Cottrell} {et~al.}(2005){Cottrell}, {Albrow}, {Barnes}, \&
  {Kershaw}}]{Cottrell2005}
{Cottrell}, P., {Albrow}, M., {Barnes}, S., \& {Kershaw}, G. 2005, {Critical
  Design Review for the Southern African Large Telescope High-Resolution
  Spectrograph}, Tech. rep.

\bibitem[{{Den Hartog} {et~al.}(2003){Den Hartog}, {Lawler}, {Sneden}, \&
  {Cowan}}]{Den2003}
{Den Hartog}, E.~A., {Lawler}, J.~E., {Sneden}, C., \& {Cowan}, J.~J. 2003,
  ApJS, 148, 543

\bibitem[{{Gilliland} {et~al.}(1998){Gilliland}, {Bono}, {Edmonds}, {Caputo},
  {Cassisi}, {Petro}, {Saha}, \& {Shara}}]{Gilliland1998}
{Gilliland}, R.~L., {Bono}, G., {Edmonds}, P.~D., {Caputo}, F., {Cassisi}, S.,
  {Petro}, L.~D., {Saha}, A., \& {Shara}, M.~M. 1998, ApJ, 507, 818

\bibitem[{{Gratton} {et~al.}(2004){Gratton}, {Sneden}, \&
  {Carretta}}]{Gratton2004}
{Gratton}, R., {Sneden}, C., \& {Carretta}, E. 2004, ARA$\&$A, 42, 385

\bibitem[{{Hannaford} {et~al.}(1982){Hannaford}, {Lowe}, {Grevesse}, {Biemont},
  \& {Whaling}}]{Hannaford1982}
{Hannaford}, P., {Lowe}, R.~M., {Grevesse}, N., {Biemont}, E., \& {Whaling}, W.
  1982, ApJ, 261, 736

\bibitem[{{Harris}(1996)}]{Harris1996}
{Harris}, W.~E. 1996, AJ, 112, 1487

\bibitem[{{Hartwick} \& {Hesser}(1974)}]{Hartwick1974}
{Hartwick}, F.~D.~A. \& {Hesser}, J.~E. 1974, in BAAS, Vol.~6, BAAS, 216

\bibitem[{{Kurucz} \& {Peytremann}(1975)}]{Kurucz1975}
{Kurucz}, R.~L. \& {Peytremann}, E. 1975, SAO Special Report, 362

\bibitem[{{Lawler} {et~al.}(2001){Lawler}, {Bonvallet}, \&
  {Sneden}}]{Lawler2001}
{Lawler}, J.~E., {Bonvallet}, G., \& {Sneden}, C. 2001, ApJ, 556, 452

\bibitem[{{Lee}(1977)}]{Lee1977}
{Lee}, S.~W. 1977, A$\&$AS, 27, 381

\bibitem[{{McWilliam} \& {Bernstein}(2007)}]{McWilliam2007_accepted}
{McWilliam}, A. \& {Bernstein}, R.~A. 2007, ArXiv e-prints, 709.1964

\bibitem[{{Nordsieck} {et~al.}(2001){Nordsieck}, {Burgh}, {Kobulnicky},
  {Williams}, {O'Donoghue}, {Percival}, \& {Smith}}]{Nordsieck2001}
{Nordsieck}, K.~H., {Burgh}, E.~B., {Kobulnicky}, H.~A., {Williams}, T.~B.,
  {O'Donoghue}, D., {Percival}, J.~W., \& {Smith}, M.~P. 2001, in BAAS,
  Vol.~33, BAAS, 1465

\bibitem[{{Sharp} {et~al.}(2006){Sharp}, {Saunders}, {Smith}, {Churilov},
  {Correll}, {Dawson}, {Farrel}, {Frost}, {Haynes}, {Heald}, {Lankshear},
  {Mayfield}, {Waller}, \& {Whittard}}]{Sharp2006}
{Sharp}, R., {Saunders}, W., {Smith}, G., {Churilov}, V., {Correll}, D.,
  {Dawson}, J., {Farrel}, T., {Frost}, G., {Haynes}, R., {Heald}, R.,
  {Lankshear}, A., {Mayfield}, D., {Waller}, L., \& {Whittard}, D. 2006, in
  Proceedings of the SPIE, Vol. 6269, Ground-based and Airborne Instrumentation
  for Astronomy, ed. I.~S. {McLean} \& I.~{Masanori}, 62690G

\bibitem[{{Sneden}(1973)}]{MOOG}
{Sneden}, C. 1973, PhD thesis, University of Texas at Austin

\bibitem[{{Wylie} {et~al.}(2006){Wylie}, {Cottrell}, {Sneden}, \&
  {Lattanzio}}]{Wylie2006}
{Wylie}, E.~C., {Cottrell}, P.~L., {Sneden}, C.~A., \& {Lattanzio}, J.~C. 2006,
  ApJ, 649, 248

\end{thebibliography}

\end{document}